# Prototyping of Open Source NB-IoT Network

Chieh-Chun Chen[*], Ray-Guang Cheng[*], Chung-Yin Ho[*], Matthieu Kanj[**],
Bruno Mongazon-Cazavet[+], Navid Nikaein[++],
[*]National Taiwan University of Science and Technology (NTUST), [**]B-COM, France,
[+]Nokia Bell Labs, France, [++]EURECOM, France
Email: crg@gapps.ntust.edu.tw; bruno.mongazon-cazavet@nokia-bell-labs.com

*Abstract*—

Narrowband Internet-of-Things (NB-IoT) is one of the major access technologies proposed to support massive machine type communications (mMTC) services for the $5^{th}$ generation (5G) mobile networks. Many emerging services and networking paradigms are expected to be developed on top of NB-IoT networks. This paper summarizes the steps required to build up an open source narrowband Internet-of-Things (NB-IoT) network. This work is a joint research and development (R&D) result from industry and academic collaboration. The open source NB-IoT enhanced Node B (eNB) is jointly developed by EURECOM, B-COM and NTUST based on the well-known OpenAirInterface[TM] (OAI) open source Long-Term Evolution (LTE) eNB. The NB-IoT eNB is successfully connected to an evolved packet core (EPC) developed by Nokia Bell Lab. We demonstrate how to use commercial off-the-shelf (COTS) NB-IoT module to forward its sensing data to the Internet via the open source NB-IoT network.

*Index Terms*—**NB-IoT, MAC, open source**

## I. INTRODUCTION

The $5^{th}$ generation (5G) network is designed to support three primary use cases of enhanced mobile broadband (eMBB), ultra-reliable and low latency (uRLLC) and massive machine type communications (mMTC). The deployment of new services in cellular networks may greatly change the way we live, work, and play. Current generations of hardware/software for radio access network (RAN) consist of many proprietary elements, which increase the cost for the operators and suppress innovation. Open source software and hardware reference design enables faster and permissionless innovation and is one of the most efficient ways to accelerate 5G/B5G product development [1]. Open source software running on general purpose processors can greatly simplify network access, reduce cost, increase flexibility, improve innovation speed and accelerate time-to-market for introducing new services [2].

OpenAirInterface[TM] (OAI) [3] is one of the open-source solutions implementing the 3rd Generation Partnership Project (3GPP) Long-Term Evolution (LTE) standard developed by EURECOM. Based on the initial work of OpenAirInterface[TM], a non-profit consortium named OpenAirInterface[TM] Software Alliance (OSA) [2] is funded to foster a community of industrial and research contributors for open source software and hardware development for the core network (EPC), radio access network (RAN) and user equipment (UE) of 5G cellular stack on commercial off-the-shelf (COTS) hardware. One of the on-going projects in OSA is narrowband Internet-of-Things (NB-IoT) enhanced Node B (eNB) [4]. NB-IoT is a 3GPP standard defined to support mMTC service. NB-IoT is evolved from LTE-Advanced (LTE-A) standard orthogonal frequency-division multiple access (OFDMA) technology but each carrier only requires 180 kHz bandwidth. It supports 'stand-alone,' 'in-band,' and 'guard band' modes. NB-IoT utilizes repetitions to enhance the coverage of the eNB. Up to three coverage enhancement (CE) levels can be supported by one eNB. The impact of repetition can be found in [5]. The OAI NB-IoT eNB project implements the physical layer functions and the protocols stacks 3GPP R'13 standard.

The general principle in implementing an NB-IoT eNB is to start with the LTE protocols and reduce them to a minimum with functionalities optimized for NB-IoT. The protocol stacks of NB-IoT are the same as LTE, which include packet data convergence protocol (PDCP), radio resource control (RRC), radio link control (RLC) and medium access control (MAC) layers. The protocol stacks are mainly contributed by NTUST [4]. The physical (PHY) layer functionalities are developed by B-COM. All downlink/uplink channels and signals were developed by considering the new requirements of NB-IoT system (e.g. repetitions, 15 kHz and 3.75 kHz subcarrier spacing, etc.). Most of the effort in implementing the OAI NB-IoT eNB was mainly on the MAC layer. It is because the repetition technique adopted by NB-IoT results in a totally different implementation of random-access channel and MAC scheduler than that of LTE. The challenge and the basic design concept of an NB-IoT MAC scheduler can be found in [6]. In addition to MAC layer, the RLC, PDCP, and RRC stacks are also revised based on the NB-IoT specification.

The MAC layer architecture of the OAI NB-IoT eNB can be found in [7]. The flow chart of main functions and the MAC scheduler and key messages required by a UE to attach an NB-IoT eNB and their scheduling interval were elaborated. The proposed OAI NB-IoT eNB further supports the functional spilt Option 6. The standard interface between MAC and PHY layers follow the standard functional application platform interface (FAPI) and network functional application platform interface (nFAPI) specification defined by Small Cell Forum [8]. Currently, we can use COTS UE to connect the proposed NB-IoT eNB, and ping a Google server via Nokia Bell Labs software EPC (a.k.a. LTEBox). This paper presents the steps and software tools required for installing an open source NB-IoT network in NTUST. Academic and industrial researchers can duplicate the open source NB-IoT network environment in their laboratories can to develop/test their own emerging services and networking paradigms. The rest of the paper is organized as below. Section II summarizes the guideline for installing NB-IoT eNB and UE. This paper focuses on the building up the NB-IoT testing environment by 3GPP R'13, and the demonstration of the data transition is given in Sec. III. Conclusion and the future work are given in Sec. IV.

## II. GUIDELINE FOR INSTALLING NB-IOT ENB & UE

### A. Environment

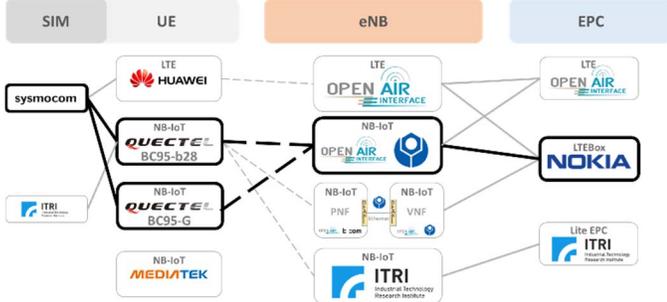

Fig. 1 The testing environment in BMW Lab.

Figure 1 shows the testing environment adopted in Broadband Multimedia Wireless Laboratory (BMW Lab.), NTUST. In this paper, we will focus on the deployment of an OAI NB-IoT network (i.e., the blocks with bold lines). We use LTEBox developed by Nokia Lab. for our EPC. For comparison, we also deploy a commercial version of NB-IoT eNB and EPC developed by Industrial Technology Research Institute (ITRI), Taiwan. Two versions of the NB-IoT eNB were deployed. The first version implements the functional-split feature of the eNB using nFAPI, where the PHY and the higher stacks are running in a physical network function (PNF) and a virtual network function (VNF), respectively. VNF and PNF are run on two personal computers (PCs) connected by Ethernet. The details of design of this version can be referred to [7]. The second version is an integrated NB-IoT eNB considered in this paper. The COTS UEs of a BC95-B28 module and a BC95-G module from Quectel are tested [9]. A configurable SIM card from sysmocom [10] was used. The eNB is installed on a PC over Ubuntu16.04 operational system with a Universal Software Radio Peripheral (USRP™) B200 or B210 software-defined radio kit [11]. The source code of NB-IoT eNB can be downloaded from OAI develop-nb-iot branch (81b8706bdeff6cc5bec8f260ad582102483eadee) [12]. We utilize Nokia LTEBox as EPC since OAI EPC does not support NB-IoT at this moment. Note that academic researchers may contact with Bruno to request for non-disclosure agreement (NDA) from Nokia Bell Labs., France, to use the LTEBox. We use BC95-B28 and BC95-G, and the frequency band is band 28.

### B. Install NB-IoT eNB

Figure 2 shows the network configurations of eNB and LTEBox used in our environment. S1-U and S1-C (or, S1_MME) represents the user and control plane connections of the S1 interface, respectively. The following steps are used to deploy the OAI NB-IoT eNB.

**STEP 1:** *Follow the commands shown in Fig. 3 to build up NB-IoT eNB.*

**STEP 2:** *Copy the configuration file. Modify the physical parameters in the configuration file as shown in Fig. 4. Set the IP addressed of the mobility management entity (MME) and eNB based on the example shown in Fig. 5.*

**STEP 3:** *Follow the commands shown in Fig. 6 to run NB-IoT eNB.*

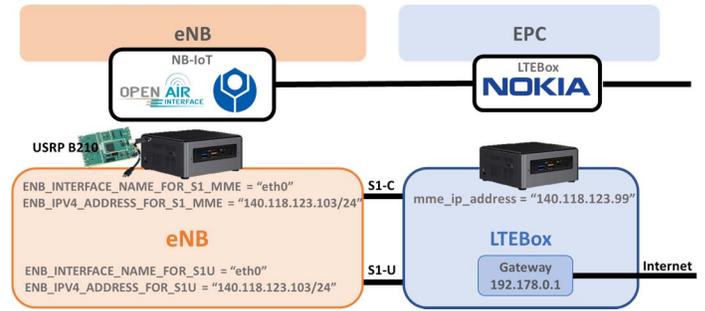

Fig. 2 The network configurations of eNB and LTEBox

- **Clone the source code & Install the packages**
$ git clone https://gitlab.eurecom.fr/oai/openairinterface5g.git
$ cd openairinterface5g; git checkout develop; cd cmake_targets
$ sudo ./build_oai -I --eNB -x --install-system-files -w USRP
- **Checkout to the develop-nb-iot branch & Build the eNB (commit: 81b8706bdeff6cc5bec8f260ad582102483eadee)**
$ cd openairinterface5g; git checkout develop-nb-iot
$ cd cmake_targets; sudo ./build_oai -w USRP -x -c –eNB

Fig. 3 Build NB-IoT eNB

- **Copy the configuration file**
$ cd openairinterface5g/targets/PROJECTS/GENERIC-LTE-EPC/CONF
$ cp enb.band28.tm1.50PRB.usrpb210.conf.conf ~/nbiot_test.conf
- **Confirm the configuration file is same as following parameters**
component_carriers =
({ …
eutra_band = 28;
dowlink_frequency = 780000000L;
uplink_frequency_offset = -55000000;
…
rach_raResponseWindowSize_NB = 8;
rach_macContentionResolutionTimer_NB = 32;
rach_preambleInitialReceivedTargetPower_NB = -90;
nprach_SubcarrierMSG3_RangeStart = "zero"
maxNumPreambleAttemptCE_NB = 3;
numRepetitionsPerPreambleAttempt = 1;
npdcch_NumRepetitions_RA = 4;
npdcch_StartSF_CSS_RA = 2;
npdcch_Offset_RA = "oneFourth";
});

Fig. 4 Physical parameters to be modified in the configuration file

Fig. 5 IP addresses setting in configuration file

- **Run the NB-IoT eNB by your config file (e.g. nbiot_test.conf)**
$ cd openairinterface5g/cmake_targets /lte_build_oai/build
$ sudo ./lte-softmodem -O ~/*nbiot_test.conf*

Fig. 6 Run NB-IoT eNB

C. *Configure the UE module*

Two types of UE modules from Quectel were tested. The BC95-B28 is the first module we used to test in [4], [7]. However, this module is not available now. The BC95G is a new version available on the market. The configurations and tools used for testing the two modules are slightly different, which are elaborated below.

a. BC95-B28
**STEP 1:** *Install UE Logviewer [9]*
**STEP 2:** *Install AT COMMAND tool (Q Navigator [9]), and set serial port parameter to Main Port (e.g. COM3) and Baudrate to 9600.*
**STEP 3:** *Run UE Logviewer to display messages at the UE side. Choose the Debug port by setting UE Port Selector=Debug Port (e.g. COM7) and Baudrate to 921600.*
**STEP 4:** *Run AT command [9] to reboot the module and config the UE module*:
**AT+NRB**
**AT+NCONFIG=AUTOCONNECT,TRUE**
**AT+NCONFIG=CR_0354_0338_SCRAMBLING,FALSE**
**AT+NCONFIG=CR_0859_SI_AVOID,FALSE**
**AT+NEARFCN=0,9448**
Note that **AT+NRB** is used to reboot the module; **AT+NCONFIG** is used to configure the UE (e.g., disable scrambling at UE side); **AT+NEARFCN** is used to lock the central frequency to 780MHz by setting EARFCN=9448.
**STEP 5:** *Sett filter conditions in UE Logviewer to display the messages received and sent by the UE module.* Some useful filter conditions that we used for Logviewer are summarized as below:
  **RRC_DEBUG_ASN**: message in RRC layer of UE side
  **NAS_DBG_NAS_MSG**: message in NAS layer of UE side
  **DCI**: message in LL1 of UE side
  **HARQ**: ACK or NACK information of messages from eNB
  **RACH**: procedure in UE side

b. BC95-G
**STEP 1:** *Install UE Monitor [9]*
**STEP 2:** *Update the firmware by Qflash [9]*
**STEP 3:** *Run UE Monitor [9] to display messages at the UE side. Open a new project to connect the Debug port=Ch B (e.g. COM14) and use message definitions from xml or fwpkg (e.g. BC95GJBR01A06.fwpkg).*
**STEP 4:** *Connect the Main port=Ch A (e.g. COM13) and send AT command [9] by UE Monitor:*
**AT+NRB**
**AT+NCONFIG=AUTOCONNECT,TRUE**
**AT+NCONFIG=CR_0354_0338_SCRAMBLING,FALSE**
**AT+NCONFIG=CR_0859_SI_AVOID,FALSE**
**AT+NCONFIG=PCO_IE_TYPE,EPCO**
**AT+NCONFIG=RELEASE_VERSION,13**
**AT+NCONFIG=MULTITONE,FLASE**
**AT+CGDCONT= 0,"IP",,,0,0,,,,,0**
**AT+NEARFCN=0,9448**
**AT+CGATT=1**
Note that **AT+CGDCONT** is used to define a PDP context parameter; **AT+CGATT** is used to attach the UE to core network; and the other settings used in BC-95G are the same as that we used in BC95-B28;.
**STEP 5:** *Adopt the same filter conditions used for BC95-B28 to display the observed messages in UE Monitor.*

D. *Configure the LTEBox EPC*

The OAI NB-IoT eNB only supports the control plane solution. Hence, the IP address of S1-UP is set to be the same as that of S1-CP.

E. *Configure the SIM card*
**STEP 1:** *Prepare a PC/SC Smart Card Reader that can be used in Linux.*
**STEP 2:** *Install the necessary software for the card reader as shown in Fig. 7. If the installation is correct, the pcsc_scan should be executed after the SIM card is inserted into the card reader, then we can see the information of card reader as illustrated in Fig. 8.*
**STEP3:** *Install the burning software (PySIM[13]) by the command shown in Fig. 9.*
**STEP4:** *Configuring the SIM card following the guideline provided from the Gitlab Wiki of openairinterface5G [14]. After burning the SIM card successfully, it will show the information as Fig. 10.*
**STEP5:** *If you want to see the SIM card information after burning, you need to checkout to master branch and run the command shown as Fig. 11.*

- **Install the card reader software**
$ sudo apt-get install pcscd pcsc-tools libccid libpcsclite-dev
$ sudo pcsc_scan  /* Can see the card reader information */

Fig. 7 Install the card reader software

Fig. 8 SIM card information

- **Install PySIM**
  $ git clone git://git.osmocom.org/pysim pysim; cd pysim
- **Install the PySIM requirements**
  $ sudo apt-get install python-pyscard python-serial python-pip
  $ sudo pip install pytlv
  $ git checkout zecke/tmp2

Fig. 9 Install PySIM

Fig. 10 Burning successfully

Fig. 11 SIM card information after burning

## III. DEMONSTRATIONS

The network configuration of NB-IoT testing is show in Fig. 12. We use two PCs to run the NB-IoT eNB and LTEBox, respectively; one PC to control UE module (BC95-B28 or BC-95G). The NB-IoT UE module initiates a random-access procedure to connect to the NB-IoT eNB and starts the attach procedure with LTEBox to authenticate the SIM card information. During the attach procedure, the NB-IoT UE module transmits the message and the NB-IoT eNB forward these messages to the LTEBox. Fig. 13 shows the S1AP messages exchanged between eNB and MME during attach procedure using Wireshark.

Fig. 12 Network configuration for NB-IoT testing

Fig. 13 S1-AP messages exchanged between eNB and MME during attach procedure using Wireshark

After completing the attach procedure, the NB-IoT UE module will get an IP address from the gateway and we use AT command (AT+NPING [9]) to ping to Google DNS server (8.8.8.8) and the results are shown in Fig. 14.

Fig. 14 Ping test from UE to 8.8.8.8

As shown in Fig. 15, we use AT command (AT+NSOST [9]) to send a UDP packet to the LTEBox. Note that the '48656c6c6f204e5455535' is the UDP data in hexadecimal format. At the LTEBox side, we need to open the UDP port to receive the data from NB-IoT UE module, as shown in Fig. 16. Fig. 17 shows the messages exchanged between the LTEBox and UE using Wireshark. The last message on Fig. 17 is the UDP packet transmitted by the UE.

Fig. 15 Send UDP data from UE by AT command

Fig. 16 Receive UDP data in LTEBox

Fig. 17 Get the UDP packet from Wireshark

## IV. CONCLUSIONS AND FUTURE WORK

This paper summarizes the steps required to build up an open-source NB-IoT network. We describe the installation and configuration procedures of UE module, NB-IoT eNB and EPC in a step-by-step manner. Currently, the uplink data transmission is completed and the server can successfully receive the UDP packet. However, we found that the transmission of a data packet longer than 4 bytes may not be successfully decoded at the PHY layer of eNB. We use hardcode to revert some specific bits to fix the bugs by setting the compiling flag of NB_IOT_CRC_REVOVERY to TRUE in cmake_targets/CMakeLists.txt. Everyone can set the flag to be FALSE for revealing the issue and help us to solve the bugs. Multiple UEs stability testing can then be conducted.

## ACKNOWLEDGMENT

This work was supported in part by the Ministry of Science and Technology, Taiwan, under Contract 108-2221-E-011-041-MY2 and is conducted under the "Flagship Program on 5G Communication Systems and Intelligent Applications" of the Institute for Information Industry which is subsidized by the Ministry of Economic Affairs of the Republic of China. This is a joint work contributed by Bing-Zhi Hsieh, Wan-Rong Tsai, Kai-Hsiang Hsu, Tian-Jen Liu, Wei-Tai Chen, Yi-Chun Lin, Jing-Wei Chen, Chang-Sheng Liu, Li-Ping Yu, Cheng-Hsun Yang, Chin-Wei Kang, and Ting-An Lin, NTUST, and Vincent Savaus, B-COM.